\documentclass[journal=cmatex,manuscript=article, layout=twocolumn]{achemso}

\usepackage{fontsize}
\changefontsize[11pt]{10pt}
\usepackage{times,amsmath}
\usepackage{epsfig}
\usepackage{color}
\usepackage{longtable}
\usepackage{ulem}
\usepackage{graphicx}
\usepackage{dcolumn}
\usepackage{bm}
\usepackage{bookmark}
\usepackage{tabularx}
\usepackage{booktabs}
\usepackage{hyperref}
\usepackage{multirow}
\usepackage{amssymb}

\usepackage[version=3]{mhchem}

\hypersetup{colorlinks=true, citecolor=blue, filecolor=blue, linkcolor=blue, urlcolor=blue}

\title{Study of Entropy-Driven Polymorphic Stability for Aspirin Using Accurate Neural Network Interatomic Potential}

\author{Shinnosuke Hattori}
\email{shinnosuke.hattori@sony.com}
\affiliation{Advanced Research Laboratory, Technology Infrastructure Center, Technology Platform, Sony Group Corporation, 4--14--1 Asahi-cho, Atsugi-shi 243--0014, Japan}
\author{Qiang Zhu}
\email{qzhu8@uncc.edu}
\affiliation{Department of Mechanical Engineering and Engineering Science, University of North Carolina at Charlotte, Charlotte, NC 28223, USA}

\abbreviations{ML, MD}

\date{\today}
\begin{document}

\maketitle

\begin{abstract}
In this study, we present a systematic computational investigation to analyze the long debated crystal stability of two well known aspirin polymorphs, labeled as Form I and Form II. Specifically, we developed a strategy to collect training configurations covering diverse interatomic interactions between representative functional groups in the aspirin crystals. Utilizing a state-of-the-art neural network interatomic potential (NNIP) model, we developed an accurate machine learning potential to simulate aspirin crystal dynamics under finite temperature conditions with $\sim$0.46 kJ/mol/molecule accuracy. Employing the trained NNIP model, we performed thermodynamic integration to assess the free energy difference between aspirin Forms I and II, accounting for the anharmonic effects in a large supercell consisting of 512 molecules. For the first time, our results convincingly demonstrated that Form I is more stable than Form II at 300 K, ranging from 0.74 to 1.83 kJ/mol/molecule, aligning with the experimental observations. Unlike the majority of previous simulations based on (quasi)harmonic approximations in a small super cell, which often found the degenerate energies between aspirin I and II, our findings underscore the importance of anharmonic effects in determining polymorphic stability ranking. Furthermore, we proposed the use of rotational degrees of freedom of methyl and ester/phenyl groups in the aspirin crystal, as characteristic motions to highlight rotational entropic contribution that favors the stability of Form I. Beyond the aspirin polymorphism, we anticipate that such entropy-driven stabilization can be broadly applicable to many other organic systems and thus our approach, suggesting our approach holds a great promise for stability studies in small molecule drug design.
\end{abstract}


\vskip 300 pt

\section{Introduction}
The stability of polymorphs, characterized by free energy differences, remains a critical issue in the study of organic crystals\cite{Nyman2015-rq, Cruz-Cabeza2015-rb}. To make reliable predictions, one aims for a lattice energy precision well below the threshhold of chemical accuracy (4 kJ/mol, about 43 meV), which is necessary to distinguish between competing polymorphs. The stability of polymorphs is essential for many applications, including the control of drug solubility \cite{sun2017microstructure}, new drug and materials development \cite{corpinot2018practical, zhu2023organic} and the advancement of computational chemistry methods \cite{ouvrard2004toward, wen2012accidental, Reilly2014-gp, leblanc2016evaluation, Vaksler2021-uv}. The identification of new polymorphs, which undergo phase selection under external environmental influences such as temperature and pressure, is crucial for understanding synthesis and deposition processes. 

\begin{figure}[htbp]
  \centering
  \includegraphics[width=\linewidth]{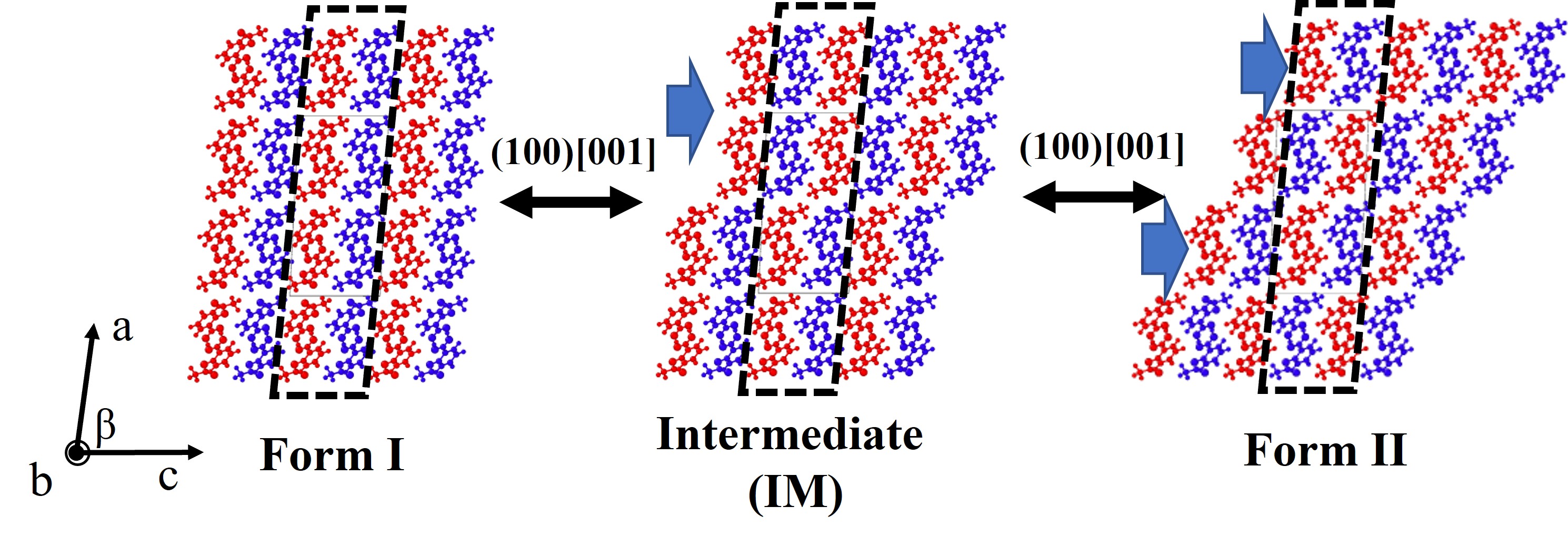}
  \caption{The schematic representations of aspirin Forms I and II, as well as the intermediate structure viewed from the projection of ac plane. For convenience, a schematic transition path between I and II via the shear deformation along the (100)[001] direction is also depicted. In all structures, molecules are colored in red and blue to distinguish their orientations.}
  \label{fig:aspmodel}
\end{figure}

In recent years, aspirin's polymorphism, have been the subject of intensive studies due to the popularity of aspirin in practical application \cite{Vishweshwar2005-dz,Bond2007-cx,Bond2007-uw,Bond2010-fg,Vaksler2021-uv, Shtukenberg2017-ar}, as well as the contrasting outcomes observed between experimental results and computational models\cite{wen2012accidental, huang2013accelerating, Reilly2014-gp}.
Form I is by far the most easily selected polymorph of aspirin crystals, whereas Form II, although structurally very similar to I, requires experimental ingenuity to synthesize. From the perspective of crystal packing, both forms exhibit a layered structure characterized by densely packed aspirin dimers at the (100) plane \cite{leblanc2016evaluation, Vaksler2021-uv}, as illustrated in Figure \ref{fig:aspmodel}. Despite their structural resemblance, the primary distinctions between two forms are how the layers are stacked and the orientation of aspirin dimer pairs within each layer. In both forms, there exist two types of molecular orientations (colored in red and blue in Figure \ref{fig:aspmodel}). In Form I, the red and blue dimers are alternatively replicated along the (001) direction, whereas Form II can be consider as a shear-slip modulation of Form I, with a lateral shift of each aspirin dimer layer along the (100)[001] direction, resulting in an alternative layer stacking arrangement. Consequently, these subtle yet non-negligible differences in the packing and orientation of aspirin dimers between Forms I and II lead to distinct thermodynamic behaviors, underscoring the complexity of polymorph stability in pharmaceutical compounds.

Despite the fact that the Form I has been well known for over a century, the successful synthesis of Form II was reported only within the past two decades \cite{ouvrard2004toward, Vishweshwar2005-dz, Bond2007-cx, Bond2010-fg, Bond2007-uw}. Such a contrast naturally raises a prevailing question in the field of aspirin research: \textit{why is Form I more commonly preferred under the standard experimental conditions?}
In recent years, multiple experiments have shown that Form II can be induced through a variety of external conditions, including the cooling rate, choice of solvent \cite{Bond2007-uw}, pressure\cite{Bond2010-fg}, indentation \cite{ghosh2012elastic}, and laser irradiation \cite{Tsuri2018-tk}.
These observations underscore the necessity for a thorough evaluation of the thermodynamic properties of aspirin crystals, pointing towards the influence of external factors on polymorph stability. Such insights are crucial for advancing our understanding of aspirin polymorphism and may have broader implications for the study of other pharmaceutical compounds.

To elucidate the observed difference in experiments, substantial computational efforts has been undertaken to quantify the energy differences between aspirin I and II \cite{wen2012accidental, huang2013accelerating, Reilly2014-gp, leblanc2016evaluation, Vaksler2021-uv, Li2022-ro}.
In these studies, the interlayer hydrogen bonding interactions of methyl groups were often considered as the main differences between Form I and II \cite{Bond2007-uw,Reilly2014-gp,Li2022-ro}. To date, most state of the art electronic structure methods (e.g., RI\_MP2 \cite{wen2012accidental}, MP2C \cite{huang2013accelerating}, PBE+MBD \cite{Reilly2014-gp}, B86bPBE-XDM \cite{leblanc2016evaluation}) found a lattice energy difference within 0.3 kJ/mol between the two polymorphs. When zero-point vibrational energies are included, the energy different decreases by 0.1 kJ/mol \cite{leblanc2016evaluation}. Given that the energy difference is comparable to a thermal vibration at room temperature, incorporating finite temperature effects becomes crucial. Indeed, a computational study based on the harmonic phonon approximation using PBE+MBD found that  Form I's stability is enhanced by a unique low-frequency vibration mode around 37 cm$^{-1}$ related to a mixture of motions of whole molecules with concerted motions of methyl groups from different molecules in and out of phase with one another \cite{Reilly2014-gp}. 
However, several more recent studies based on different methods have reported nearly indistinguishable free energy differences between two polymorphs, after considering harmonic, quasiharmonic or anharmonic phonon contributions on a small super cell \cite{leblanc2016evaluation, Li2022-ro}. Even with the same PBE+MBD functional, it was found that both forms I and II possess similar vibrational modes around 35 cm$^{-1}$ \cite{Raimbault2019-us}. On the other hand, the experimental evidence supporting the existence of low-frequency phonon modes in both forms remains contentious \cite{laman2008narrow, Li2022-ro}. The disparity naturally leads to a speculation that Form I is favored due to other factors. 

Considering an organic crystal at the room temperature condition, it is reasonable to expect that the anharmonic effects may become pronounced in determining the polymorph's free energy ranking. A standard approach to evaluate the free energy difference of two solid polymorph is thermodynamic integration (TI) \cite{dfrenkel96:mc} based on molecular dynamics (MD) simulation in conjunction with the DFT method \cite{haskins2017finite}. Although such a DFT-MD method has been employed to predict aspirin's Raman spectrum within a small supercell \cite{Raimbault2019-us}, the cost of DFT-MD calculations on the large supercell remains heavy and limits the scope of comprehensive thermodynamic analysis of aspirin crystals, which cannot be surpassed by similar approaches, especially for free energy evaluations.


Encouragingly, recent development of neural network interatomic potentials (NNIPs), has achieved a significant progress in balancing accuracy and computational efficiency \cite{Zeng2023-nf,Batzner2022-je,Musaelian2023-gs, Deng2023-xx}. These NNIPs handle E(3) equivariants and enhance accuracy even when training data are scare, by preserving identity through symmetric operations in 3D space. Among them, the Allegro package has demonstrated remarkable scalability and successfully simulated large systems with over 100,000 atoms with the aid of large GPU arrays\cite{Kozinsky2023-iv,Ibayashi2023-ya}. This breakthrough encourage us to explore the use of NNIPs to study the stability of aspirin polymorphs via high-fidelity MD simulations. Our initial step involved the development of accurate interatomic potentials tailored for aspirin polymorphs. Following this, we utilized Allegro to train an NNIP model to achieve an optimal balance between simulation accuracy and computational efficiency for organic crystals.
Finally, we employed TI techniques to calculate the free energy differences between Forms I and II and quantified the contributions that are related anharmonic effects.
As elucidated in subsequent sections, our findings reveal that the developed framework is not only effective for this specific study but also offers a generalizable approach for the analysis of other organic crystals. 


\section{Computational Methods}
\subsection{NNIP Model Training and Validation}

To obtain an accurate NNIP model, we began with preparing three sets of training data as follows. First, we performed a series of MD simulations using the general-purpose Allegro model that were pre-trained on the SPICE \cite{Eastman2023-lx} dataset for the known aspirin forms (including I , II and another recently identified metastable Form IV \cite{Shtukenberg2017-ar}) at 300 K for a duration of 10 ps. From these MD trajectories, 1200 representative structures (400 structures per polymorph) were selected to from the dataset A that represents three most important energy basin around the experimentally identified polymorphs. Second, to explore more crystal configurations that are inaccessible through direct room temperature MD simulations of the known aspirin polymorphs, we developed a automated computational pipeline to harness low-energy hypothetical crystal structures using the PyXtal code \cite{Fredericks2021-rz} based on an evolutionary crystal structure prediction (CSP) algorithm \cite{Lyakhov-CPC-2013, QZhu-Acta-2012, QZhu-JACS-2016}. We note that a similar workflow based on PyXtal has been demonstrated in a recent study on GaN \cite{Santos-Florez2023bending}, and the use of PyXtal was also integrated for automated NNIP training of inorganic materials \cite{Menon2024-pk} and crystal structure prediction \cite{janmohamed2024multi}. In this work, we performed a systematic crystal packing search on 10 common space groups with one aspirin molecule in the asymmetric unit ($Z'$=1) within 200 generations and 100 structures for each generation, based on the classical General Amber Force Field (GAFF) \cite{gaff}. Furthermore, we removed the duplicate structures from the identified CSP structures and refined their energy ranking using the ANI-2x potential\cite{Devereux2020-fn}, resulting in 4387 low-energy CSP structures in Dataset B to represent a diverse set of energy minima. Finally, we also chosen 30 lowest-energy hypothetical CSP structures to run 5 ps MD simulations and sampled a total of 1200 configurations (40 configurations per MD simulation). These configurations were labelled as dataset C to represent 30 low-energy basins in the potential energy surface of the aspirin crystal.

In parallel, we performed additional MD simulation at 350 K for the intermediate structures (IM) as depicted in Figure \ref{fig:aspmodel} at 350 K. This results in 200 configurations for dataset D that aims to explore the transition states between aspirin I and II. Unlike the aforementioned datasets, this dataset won't be seen during training, but will be used in the stage of validation only. 

\begin{figure}[!htbp]
  \centering
  \vspace{-2mm}
  \includegraphics[width=\linewidth]{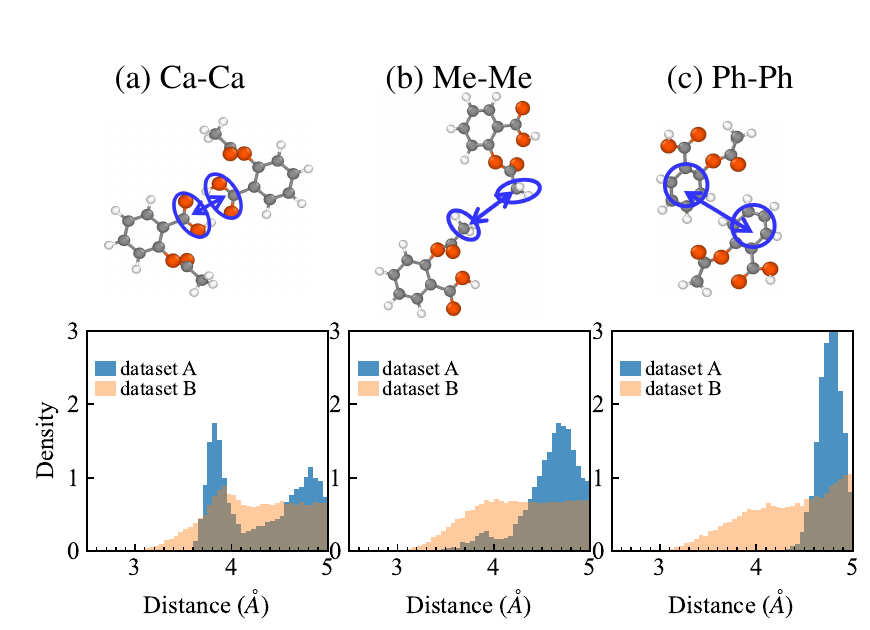}
  \caption{Representative intermolecular contacts between functional groups, including (a) carboxylic acid - carboxylic acid, b: methyl - methyl, and (c) phenyl - phenyl. The bottom panels show the distance distributions for each contact for dataset A (blue) from MD simulation and B (orange) from crystal structure prediction.}
  \label{fig:coverage}
\end{figure}

It is well known that the quality of machine learning model crucially depends on representation capability of training data. In computation chemistry, the primary molecular interactions in a crystal are generally believed to be contributed by the interactions between the representative functional groups. In the case of aspirin crystals, the most important functional groups are carboxylic acid (Ca), methyl group (Me) and phenyl group (Ph). Therefore, we proposed the use of distance distributions between functions groups within the aspirin crystal to infer the coverage of our training data. Figure \ref{fig:coverage} displayed the comparison of distributions of three most important functional interactions (including Ca-Ca, Me-Me and Ph-Ph) from both datasets A and B. Clearly, Figure \ref{fig:coverage} suggests that dataset A, derived from MD simulations, has a narrower distribution peak than the CSP dataset B. This quantitative comparison demonstrates that it is necessary to use a more efficient structure sampling approach prior to the NNIP model training. 

For each configuration in the collected datasets, we performed single point energy calculation to obtain the reference energy, forces and stress tensors, with the combination r2SCAN functional and D4 correction for the van der Waals (vdW) correction \cite{Ehlert2021-hd}. Within the framework of the Allegro code \cite{Musaelian2023-gs}, we trained the NNIP models with different choices of training data and strategies (see more details about the choices of hyperparameters in the appendix). 
As shown in Table \ref{table:nnip}, we found that 
the model integrating datasets A, B, and C, and pre-trained model with SPICE, delivered the most favorable outcomes, achieving the root mean square errors (RMSEs) of 0.12 kJ/mol/molecule for energy, 3.92 kJ/mol/\AA ~for force, and 0.06 GPa for stress tensors within the validation dataset D. On the other hand, the model with dataset A without pre-trained has the RMSE values of 0.39 kJ/mol/molecule for energy, 5.27 kJ/mol/\AA ~for force, and 0.03 GPa for stress tensors. Our results demonstrate that developing a tailored NNIP for aspirin only can achieve a considerably better accuracy as compared to the commonly used general-purpose force field.   

\begin{figure}[htbp]
    \centering    
    \includegraphics[width=0.9\linewidth]{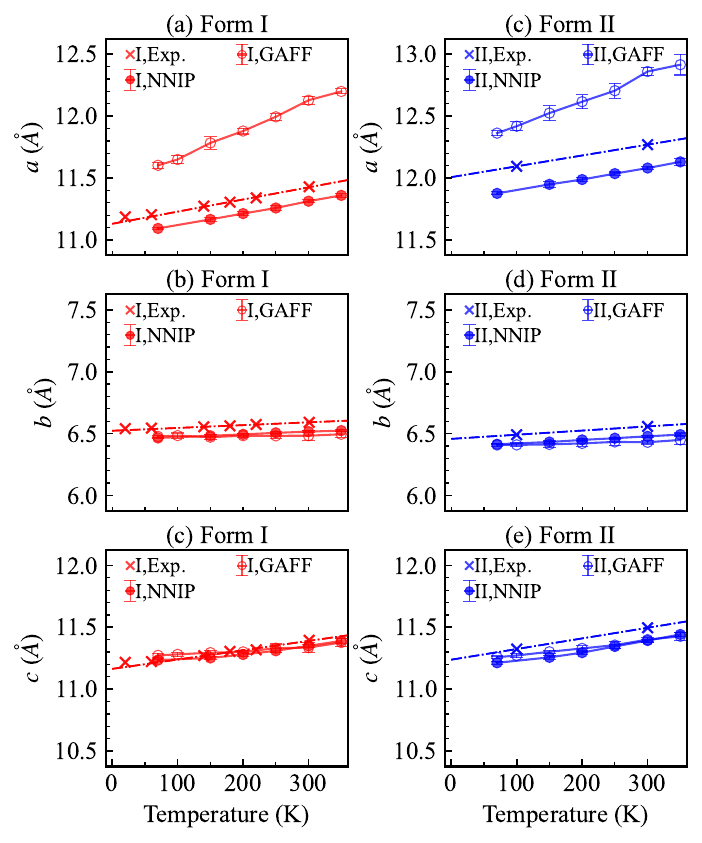}
    \caption{The comparison of lattice constants of aspirin crystals between experimental data (cross markers), NNIP (solid circles) and GAFF (open circles) results for Forms I (the upper panel from a to c) and II (the lower panel from d to f) sampled from a 100 ns MD simulation of the 4$\times$4$\times$4 supercell structures. The experimental data were taken from CCDC, including ACSALA01\cite{Kim1985-qw}, ACSALA03, ACSALA04, ACSALA05, ASCALA06 and ACSALA08\cite{Wilson2002-vx} for Form I, and ACSALA13\cite{Vishweshwar2005-dz} and ACSALA17\cite{Chan2010-xc} for Form II.}
    \vspace{-3mm}
\label{fig:cell}
\end{figure}

\begin{table*}[htbp]
  \centering
  \footnotesize
  \begin{tabularx}{380pt}{c|ccccccccccccccc}
    \hline\hline
        & \multicolumn{4}{c}{ Energy RMSE} & & \multicolumn{4}{c}{ Force RMSE } &  & \multicolumn{4}{c}{ Stress RMSE}  \\
        & \multicolumn{4}{c}{(kJ/mol/molecule) } & & \multicolumn{4}{c}{ (kJ/mol/\AA) } &  & \multicolumn{4}{c}{(GPa) }                                                                                                                               \\
         & A   & B     & C     & D   && A    & B     & C     & D    && A    & B   & C    & D  \\\hline
(i) & 0.12 & 8.23 & 3.27 & 0.39 && 3.00 & 25.28 & 11.11 & 5.27 && 0.02 & 0.19 & 0.13 & 0.03 \\
(ii) & 0.10 & 25.12 & 3.03 & 0.26 && 2.75 & 24.02 & 90.91 & 4.63 && 0.02 & 0.31 & 0.10 & 0.03 \\
(iii)& 0.96& 1.74 & 0.63 & 0.29 && 4.18 & 4.19 & 2.39 & 4.19 && 0.08 & 0.09 & 0.07 & 0.08 \\
(iv) & 0.67& 1.46 & 0.81 & 0.12 && 4.02 & 3.89 & 2.41 & 3.92 && 0.07 & 0.08 & 0.06 & 0.06 \\

    \hline\hline
  \end{tabularx}
  \caption{The accuracy of NNIP model on different datasets and pre-training strategies, including (i) training on dataset A only, (ii) training on dataset A after pre-training on the SPICE dataset, (iii) training on dataset A, B and C, (iv) training on dataset A, B and C after pre-training on the SPICE dataset.}
  \label{table:nnip}
\end{table*}
To further clarify the accuracy of our NNIP model, we performed two additional validations. First, we tested the temperature dependence of cell parameters for both I and II by running a series of 4$\times$4$\times$4 supercell NPT MD simulations lasting for 100 ps at a range of temperatures from 70 to 350 K. In each simulation, the averaged cell parameters for the last 10 ps were used to determine the equilibrium geometry at the given temperature. For the purpose of comparison, we repeated the same procedures to obtain the temperature-dependent lattice parameters by using the classical GAFF model. The results of NNIP, GAFF and available experimental results are shown in Figure \ref{fig:cell}. As compared to the GAFF model, the NNIP results tend to achieve a better agreement with the experimental data. Importantly, in the direction which has more sensitive temperature dependence (see Figure \ref{fig:cell}a and d), the GAFF model significantly overestimates the thermal expansion, while NNIP remains to yield accurate results as compared to the experimental values. 

Second, we tested if the NNIP model can faithfully reproduce the transition pathway between aspirin I and II, which is crucial for the subsequent free energy calculation. 
Following the recent works\cite{leblanc2016evaluation, Vaksler2021-uv}, we constructed the transition path by applying a constant strain rate (0.1 \AA/ps) on the xz component of a $2\times2\times2$ supercell of Form I in a MD simulation with the NNIP model as implemented in the LAMMPS code \cite{lammps}. From this MD simulation trajectory, we selected the snapshots every 50 ps as initial structures to run subsequent calculations for sampling their thermodynamic quantities. For each snapshot structure, we performed MD simulation with two stages. First, we performed a constrained NPT MD simulation by fixing the strain of $\epsilon_\text{xz}$ for 10 ps to obtain a well relaxed configuration at the given $\epsilon_\text{xz}$ condition. Then, another 10 ps NVT MD calculations was performed to gain the averaged thermodynamic quantities. In addition, the same calculations were repeated using GAFF and DFT-MD simulations for a comparison.


\begin{figure}[htbp]
  \centering   
  \includegraphics[width=0.90\linewidth]{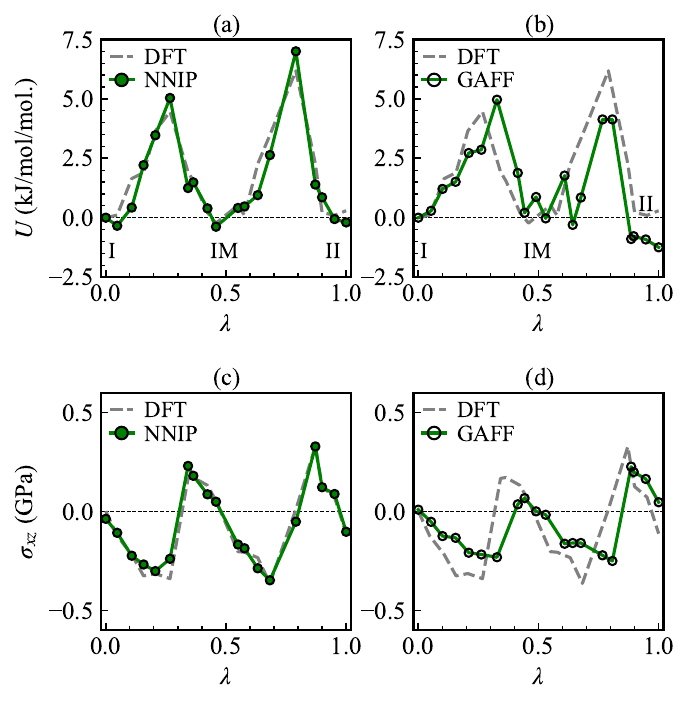}
    \vspace{-1mm}
    \caption{Computed potential energy (a-b) and $\sigma_\text{xz}$ stress values (c-d) between aspirin I-II transition. In all subplots, we used a normalized variable $\lambda$ based on $\Delta \cos \beta$ along the transition path to represent the x-coordinates. For NNIP and GAFF, the DFT results are superimposed with dashed lines for comparison. All values are averaged from the results of MD simulations with 2$\times$2$\times$2 supercells at 300 K.}
    \vspace{-2mm}
    \label{fig:valid}
\end{figure}

Figure \ref{fig:valid} displays the NNIP, GAFF and DFT results regarding the profiles of energy ($U$) and $\sigma_\text{xz}$ stress along the transition path from I to II. Clearly, both $U$ and $\sigma_\text{xz}$ profiles generated by NNIP accurately reproduce the DFT results (see Figure \ref{fig:valid}a and c). As a comparison, the GAFF also captures the general trend, however, the errors are notably larger than the NNIP results. These validations guarantee the use of NNIP in studying the free energy difference between Forms I and II within the framework of TI to be discussed in the following section.

\subsection{Free Energy Difference Calculation}
In order to determine the free energy difference between aspirin I and II, we employed the thermodynamic integration approach in conjunction with MD simulations\cite{Menon2021-an} by following a recent work on NiTi alloys\cite{Tang2022-sg}. In the context of TI, we chosen $\cos \beta$ as the integrating variable ($\lambda$) and computed the Helmholtz free energy difference $\Delta F_{\text{I} \rightarrow \text{II}}$ via integrating the transition path from I to II as follows. 
\begin{equation}
  \Delta G_{\text{I} \rightarrow \text{II}} \approx \sum_{k=0}^{n} V_k \langle\boldsymbol{\sigma}_k\rangle  \boldsymbol{\Omega}_k^{-T}: \frac{\partial\boldsymbol{\Omega}_k}{\partial\lambda_k} \Delta \lambda_k \\
\end{equation}
where $n$ is the number of sampling points along the $\lambda$-path, and $\boldsymbol{\Omega_k}$, $V_k$ and $\sigma_k$ denote the cell matrix, the cell volume and stress tensor in each sampled state. The derivative of $\boldsymbol{\Omega_k}$ can be obtained by numerical derivation of fitting function along the $\lambda$-path.
And $\Delta \lambda_{k}$ indicates the difference in $\lambda$ between sampling points $k$-1 and $k$. Furthermore, the Gibbs free energy difference $\Delta G_{\text{I} \rightarrow \text{II}}$ can be approximated as $\Delta F_{\text{I}\rightarrow\text{II}}$ by omitting the variation of $pV$ term due to its negligible contribution.




In this work, we tried TI calculations with a series of supercell models up to  2$\times$4$\times$16 (i.e. at most 512 molecules). In particular, we choose to 
focus on varying the replication on the $c$-axis (up to 185.1 \AA) in order to minimize the fluctuation of stress applied in the (100)[001] direction $(\sigma_\text{xz}$) during the sampling of high energy intermediate configuration, thus reducing the error in the free energy calculation (see details in the appendix A3).

\section{Results and Discussions}
\subsection{Free energy calculation}

\begin{figure}[htbp]
\centering
\vspace{-1mm}
\includegraphics[width=0.90\linewidth]{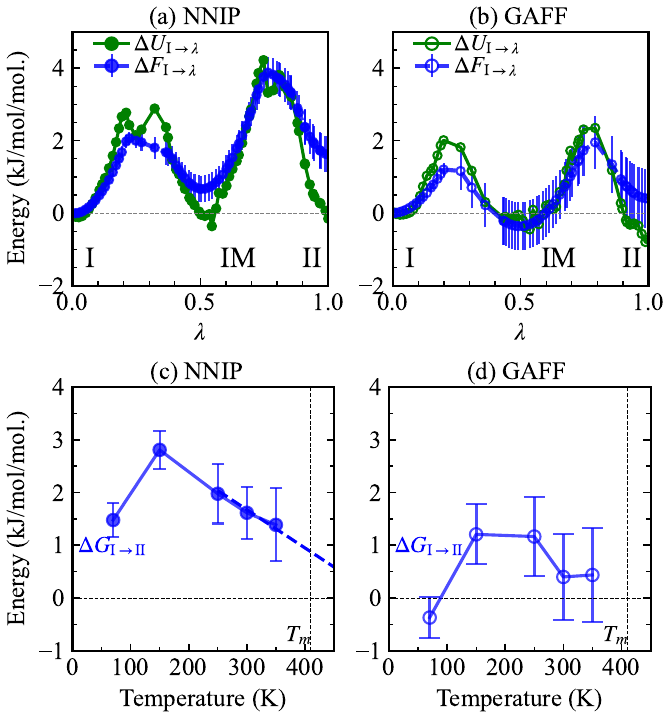}
  \caption{The simulated energy profiles along the path from Form I to II at 300 K, with (a) NNIP and (b) GAFF. (c) and (d) display the temperature dependence of the Gibbs free energy difference between Form I and II (red). Error bars in each subplot indicate the accumulated integrated error throughout the analysis. The vertical black dotted line indicates the melting point of Form I.}
  \label{fig:TI}
  \vspace{-1mm}
\end{figure}

Figure \ref{fig:TI} summarizes the free energy differences from the TI calculations between Forms I and II using the NNIP and GAFF models. As shown in Figure \ref{fig:TI}a, the TI simulations using NNIP model at 300 K suggest that Form I has a lower free energy than Form II ($\Delta G_{\text{I} \rightarrow \text{II}}$ = 1.83 kJ/mol/molecule), despite the nearly zero difference in internal energy ($\Delta U_{\text{I} \rightarrow \text{II}}$). In addition, we considered the errors of $\Delta G_{\text{I} \rightarrow \text{II}}$ in the TI simulation due to the fluctuation of stress tensors. In a $2\times4\times16$ supercell at 300 K, the estimated error is about 0.46 kJ/mol/molecule, which is notably smaller than $\Delta G_{\text{I} \rightarrow \text{II}}$, thus confirming that Form I should be more stable regardless of numerical errors. Furthermore, Figure \ref{fig:TI}c plots the temperature dependence of Forms I and II from 0 to 400 K. In the entire temperature range, it shows that $\Delta G_{\text{I}\rightarrow\text{II}}$ first increases to 2.9 kJ/mol/molecule and then gradually decreases from 150 K to 400 K. Hence, $\Delta G_{\text{I}\rightarrow\text{II}}$ remains positive in the entire temperature range from 0 to 400 K.

For a comparison, the GAFF simulations, as shown in Figure \ref{fig:TI}b and d, reveal a general trend very similar to the NNIP results. However, the accompanying errors are considerably larger. Thus, it is clear that a quantum-accuracy model is necessary when the free energy difference is near the sub kJ/mol range.

\begin{figure}[htbp]
  \centering
    \includegraphics[width=0.50\linewidth]{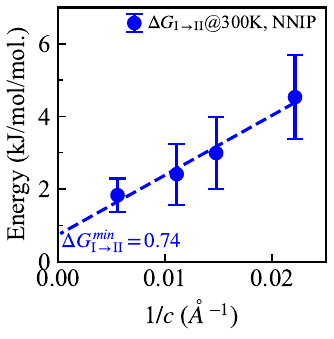}
  \caption{The computed $\Delta G_{\text{I}\rightarrow\text{II}}$ at 300 K as a function of model size (1/$c$).}
  \label{fig:size}
\end{figure}

Given that the TI results may strongly depend on the setup of simulated model size, it is necessary to further investigate the size dependence of our TI simulation. Therefore, we compared TI results by considering a variety of supercell models, as detailed in Table \ref{tab:energy_table}. Clearly, the Gibbs free energy difference, $\Delta G_{\text{I}\rightarrow\text{II}}$, decreases with the model size (in particular, it depends on the length of $c$-axis). Fitting the trend of the decrease with the reciprocal of $c$ value, we find the lower bound of $\Delta G_{\text{I}\rightarrow\text{II}}^\text{min}$ should be around 0.74 kJ/mol/molecule even when extrapolated to infinity, as shown in Figure \ref{fig:size}. In addition, the integral error of the TI method, evaluated based on the variance of the pressure at each sampling point 
is sufficiently reduced when the supercell model increases. 
Hence, it is safe to conclude that Form I is consistently stable than Form II at 300 K regardless of the supercell size choices.

Experimentally, the free energy difference between Form I and Form II crystalline polymorphs can be estimated from an empirical approach by considering the a melting point difference (10 K) between I and II \cite{Mittal2016-ei}. Assuming that (i) the free energy difference of two polymorphs at the melting point is zero, (ii) the temperature dependence of the free energy difference is linear between the melting points, and (iii) the molar heat capacity difference ($\Delta C_p$) is constant, the free energy difference ($\Delta G$) can be approximated using the equation $\Delta G \approx \Delta S_f \times \Delta T$, where $\Delta S_f$ is the melting entropy (entropy change associated with melting; fusion entropy) and $\Delta T$ is the difference in melting points (10 K). Using a typical melting entropy value for organic compounds of 56.5 J/mol-K \cite{Gilbert1999-aw}, the calculated $\Delta G$ is approximately 0.56 kJ/mol. This value is comparable to the 1 kJ/mol obtained by extrapolating the data in Figure \ref{fig:TI}c to around 410 K, which is near the experimental melting point of aspirin I. When considering the size effect, the calculated result is not significantly different from the figure value. Therefore, we conclude that our free energy difference results from TI simulation is consistent with the previous experimental reports \cite{Bond2007-cx, Mittal2016-ei}. 


\begin{table}[htbp]
\footnotesize
\caption{The computed free energy differences (kJ/mol/molecule) between aspirin forms using NNIP at 300K. Errors in parentheses are shown in RMSE.}
\label{tab:energy_table}
\centering
\begin{tabular}{ccccc}
\hline\hline
Supercell & $2\times4\times4$ & $2\times4\times6$ & $2\times4\times8$ & $2\times4\times16$ \\
\hline
$\Delta G_{\text{I} \rightarrow \text{II}}$ & 4.53 (1.16) & 2.99 (0.99) & 2.41 (0.84) & 1.83 (0.46) \\

$\Delta G_{\text{I} \rightarrow \text{IM}}$ & 2.60 (0.68) & 1.59 (0.63) & 1.49 (0.58) & 1.01 (0.34) \\
\hline\hline
\end{tabular}
\end{table}

\begin{figure}[htbp]
  \centering
  \includegraphics[width=0.9\linewidth]{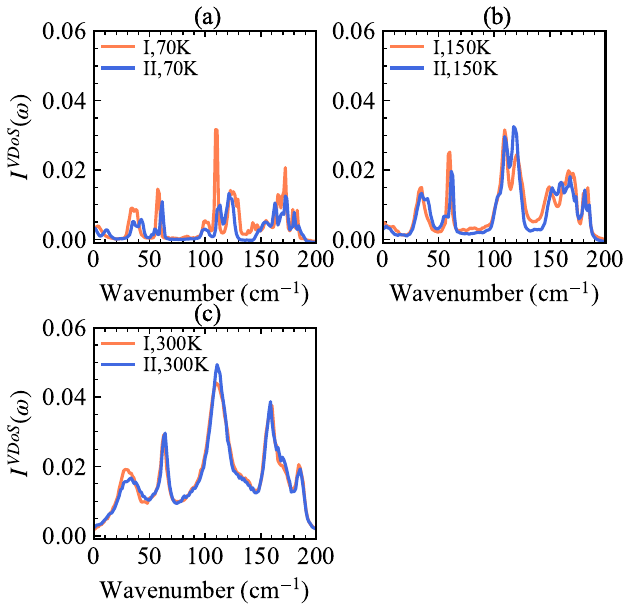}
  \vspace{-1mm}
  \caption{Vibrational density of states from NNIP-MD with (a) 70, (b) 150, and (c) 300 K. Line color shows Form I (orange) and Form II (blue).}
  \label{fig:VDOS} 
\end{figure}

As discussed earlier, the majority of previous (free) energy calculations \cite{leblanc2016evaluation, Li2022-ro}, based on the either harmonic or quasihamonic phonon approach, indicate that aspirin I and II are nearly indistinguishable under the room temperature conditions. Indeed, we also found that the vibrational density of states (VDOS) of I and II at 70, 150, and 300 K (see Figure \ref{fig:VDOS}) are nearly identical by using the Fourier method from the NNIP-MD simulation. Using the harmonic phonon approximation, the free energy difference between I and II is only 0.03 kJ/mol/molecule at 300 K. 
However, our TI results based on both the classical GAFF and more accurate NNIP models yield a consistent stability ranking as compared to the experiment. Furthermore, both GAFF and NNIP results (see Figure \ref{fig:TI}b and d) suggest that $\Delta G_{\text{I}\rightarrow\text{II}}$ first increases while the temperature rises from 0 to 150 K and then has a trend to decrease while further increasing the temperature from 150 K to a high temperature limit around 400 K. This is an indication that some anharmonic molecular motions, which cannot be captured by the harmonic phonon approximation, may have different temperature dependencies in Form I and II. To understand the physical origin of free energy difference between I and II, we proceeded to analyze the representative molecular motions from MD simulation trajectories. As discussed earlier, the majority of molecular interactions in the aspirin crystals may be characterized by the motions of functional groups. Hence, we performed NPT-MD simulations for both I and II at a variety of temperatures to track the motions of representative functions groups. Among them, two characteristic motions were identified and shall be discussed below.

\begin{figure}[htbp]
  \centering
  \includegraphics[width=\linewidth]{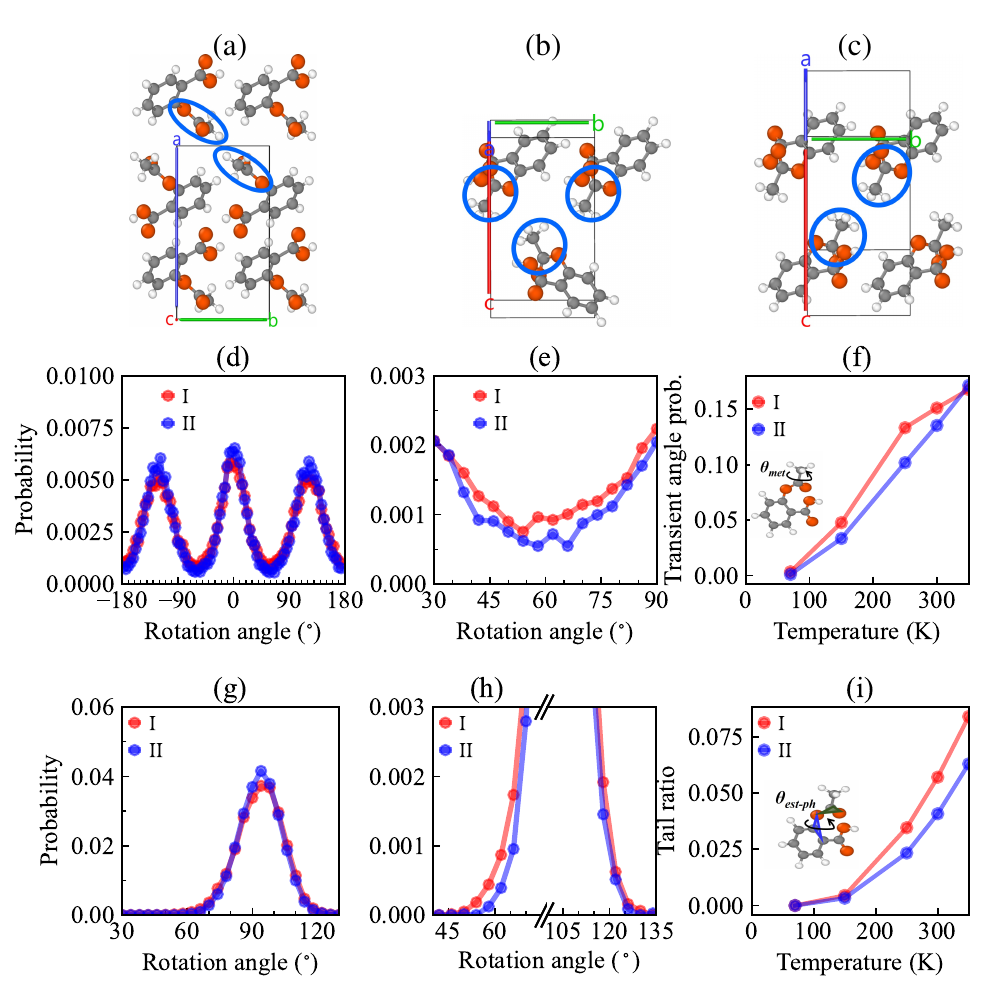}
  \caption{Two representative molecular rotations and their occurring frequencies in Form I and II from NNIP-MD simulation. (a) depicts the rotation of methyl groups viewed from the $ab$ plane in Form I, (b) and (c) highlight the rotations of ester/phenyl groups viewed from the $bc$ plane in Form I and II, respectively. For the methyl group, (d) shows the distribution of rotation angles for methyl groups evaluated with at 300K, (e) shows the low probability (transition) regions centering around 60 degree, and (f) shows the transient probability as a function of temperature. For the ester/phenyl rotations, (g) shows the distribution of rotation angle from NNIP simulation at 300 K, (g) shows the low probability regions on both left and right tails, and (i) plots low transient probability as a function of temperature.}
  \label{fig:dihed_nnip}
\end{figure}

The first motion is rotation of methyl groups. It can be seen that the methyl groups are closer together and more likely to interact in Form II in Figure \ref{fig:dihed_nnip}c than in Form I in Figure \ref{fig:dihed_nnip}b. The middle panel of Figure \ref{fig:dihed_nnip} plot the histogram of methyl group's orientation at the whole range in Figure \ref{fig:dihed_nnip}d based on the NNIP-MD results. From Figure \ref{fig:dihed_nnip}d, it is clear that the methyl group has three preferred orientation distributions centered around 0, 120 or -120 degrees, which correspond to three degenerate low-energy configurations to attach the methyl group to the backbone framework of the aspirin molecule. In the majority of MD simulations, the methyl groups rotate around within the three distinct energy basins. However, each methyl group also has the chance to switch from one basin to another via crossing the transient states, due to a small rotation barrier. Figure \ref{fig:dihed_nnip}e tracks the rotational crossover events occurring on one typical transient state around 60 degree. Clearly, such events occur more often in aspirin I as compared to aspirin II, indicating that I gains more rotational entropy than II. Figure \ref{fig:dihed_nnip}f plots the probability of observing the transient states in the MD simulations under different temperature conditions. Transition states are defined as the range of  60$\pm$30, -60$\pm$ and 180$\pm$30 degrees, and the ratio of transition states to the total is indicated. There is a clear trend that Form I generally has more rotational freedoms at low temperatures, indicating that it has a smaller activation barrier. Our results qualitatively agree with the an activation temperature range of 120 to 275 K as found in the experimental solid-state \ce{^2H}-NMR study \cite{Kitchin1999-qc}.
When the temperature becomes sufficiently high, such rotation modes become equally frequent in both Form I and II, and thus the free energy difference between I and II becomes indistinguishable as shown in Figure \ref{fig:TI}b.

Figure \ref{fig:dihed_nnip}g-i presented a similar analysis on the relative rotations between the ester and phenyl groups. We found that such rotations are mainly restricted to 90$\pm$30 degrees range in both I and II. However, Form II has a sharper peak, indicating a lower cross-over probability (i.e., a lower rotational entropy). On the other hand, the distribution on Form I has a longer tail, indicating a higher level of rotational activity, as shown in Figure \ref{fig:dihed_nnip}h. Furthermore, Figure \ref{fig:dihed_nnip}i shows that the percentages of crossover events in both forms as a function of temperature, suggesting that Form I has a greater rotational entropy than Form II at the low temperature region. 

\begin{figure}[htbp]
  \centering
  \includegraphics[width=\linewidth]{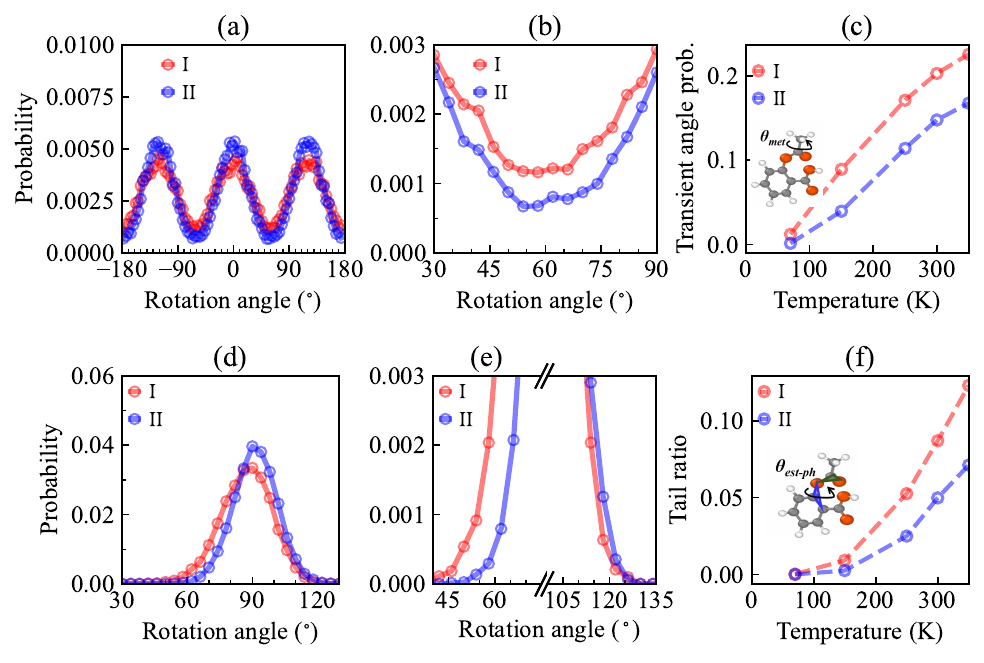}
  \caption{The statistics of molecular rotations derived from the GAFF model. For the methyl group, (a) shows the distribution of rotation angles for methyl groups evaluated with at 300K, (b) shows the low probability (transition) regions centering around 60 degree, and (c) shows the transient probability as a function of temperature. For the ester/phenyl rotations, (d) shows the distribution of rotation angle from GAFF simulation at 300 K, (d) shows the low probability regions on both left and right tails, and (f) plots low transient probability as a function of temperature.}
  \label{fig:dihed_gaff}
\end{figure}

Additionally, we repeated the analysis on the GAFF-MD results and observed a similar trend as shown in Figure \ref{fig:dihed_gaff}, confirming the robustness of our analysis. Given the above motion analysis, we conclude that cross-over molecular rotations of functional groups in Form I has a lower activation barrier as compared to that in II. Importantly, such crossover events can not be well treated within the framework of harmonic approximation. This may explain why the previous studies fail to find that I has a lower free energy by considering harmonic vibrations only.

\begin{figure}[htbp]
  \centering
  \includegraphics[width=0.90\linewidth]{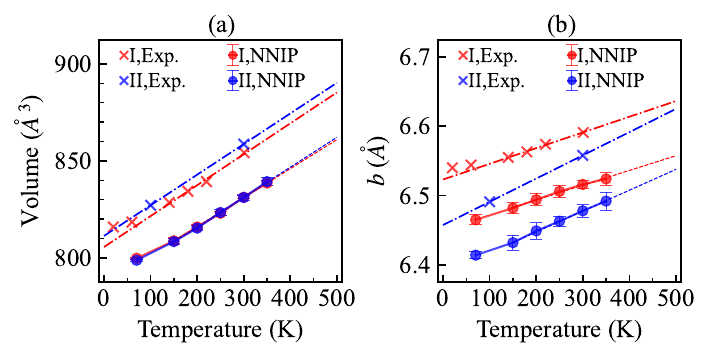}
  \caption{Temperature dependencies of unit cell volumes and $b$-axis length in both I and II. }
  \label{fig:discussion}
\end{figure}

When the temperature becomes sufficiently high and such rotation modes become equally probable for both I and II, the rotational entropy difference becomes negligible. Under such conditions, the free energy difference is expected to converge to zero. This phenomenon can also be understood from the structural aspect. As shown in Figure \ref{fig:aspmodel}, the structural difference between I and II mainly lies in the slips in the $ac$ plane, while $b$-axis is always perpendicular to the $ac$ plane and invariant to the choice of basis vectors. Figure \ref{fig:discussion} plots the evolution of cell volumes and $b$ values as a function of temperatures for both experimental and simulated data, respectively. It can be seen that the volumes (see Figure \ref{fig:discussion}a) for both I and II are nearly identical. However, the difference in $b$ is clearly more distinct. Thus we can counted $b$ as a main characteristic variable to distinguish I and II. Indeed, Figure \ref{fig:discussion}b found that the II's cell parameter $b$ is consistently lower than that of I. This indicates that a larger $b$ value in I is clearly more beneficial to promote the rotation of methyl and ester groups, thus leading to a higher entropy. With the increase of temperature, the $b$ value of II expands more rapidly than I and eventually leads to the closure of the free energy gap.

Finally, the comparative analysis between NNIP and GAFF highlights the unique value of both approaches. While the GAFF model yields less accurate energetic descriptions, it can still predict the right trend for both $\Delta G$ values (Figure \ref{fig:TI}b and d) and low-probability molecular rotation events (Figure \ref{fig:dihed_gaff}). Therefore, the generic force field models such as GAFF remains valuable due to its easy availability and low computational cost. For example, they can be used for early stage studies for the screening purposes, or the cases where a truly large scale modelling is necessary. However, in more challenging cases (e.g., the examination of polymorphic stability ranking in this study) when both a fine accuracy (e.g., $<$2 kJ/mol) and a large model size (e.g., $>$100 molecules) are critical, developing a quantum-accurate NNIPs model becomes necessary.

\section{Conclusion}
In sum, we developed a machine-learning driven interatomic potential to precisely model intermolecular interactions within aspirin crystals. Utilizing this model to analyze the polymorphic stability ranking of two aspirin crystal forms, we demonstrate that this approach yields more accurate predictions than traditional models in describing the thermal and kinetic profiles for the known aspirin crystals. By employing thermodynamic integration method that considers anharmonic effects, we determined that Form I consistently exhibits lower Gibbs free energy under the finite temperatures. Furthermore, we found that the Gibbs free energy between the two forms stems from different temperature responses of rotational motions in both I and II. For the first time, our work successfully resolves the discrepancy between the previous experimental observation and modelling results regarding the polymorph stability ranking between aspirin I and II. The success of our findings do not only highlight the importance of anharmonic effects in free energy calculation for organic crystals, but also demonstrates the power of high-fidelity NNIP models in boosting the accurate atomistic modelling for polymorphic stability of molecular crystals they may find many applications such as drug and materials design in the future.

\begin{acknowledgement}
  Computational resource of AI Bridging Cloud Infrastructure (ABCI) provided by National Institute of Advanced Industrial Science and Technology (AIST) was used. QZ thanks Hao Tang from MIT for the insightful discussion about the technical details of TI approach. 
\end{acknowledgement}


\appendix
\setcounter{equation}{0}\renewcommand\theequation{A\arabic{equation}} 
\section{Appendix}
\subsection{A1. DFT and MD simulations}\label{a1}
All DFT-MD simulations in this study were executed using the VASP code\cite{Kresse1996-vx}. MD simulations with NNIP and GAFF models in this study were executed using the LAMMMPS code\cite{lammps}.
In all MD simulations, the time step $\Delta$t was set to 0.5 fs.
In the DFT-MD simulations, the electronic states were calculated by the projector augmented wave method within the framework of density functional theory (DFT)\cite{Kresse1999-zi}.
The r2SCAN, meta generalized gradient approximation was employed for the exchange correlation energy\cite{Ehlert2021-hd,Kothakonda2022-ak,Kingsbury2022-iv}.
The empirical correction of the van der Waals interaction by the DFT-D4 approach was employed\cite{Caldeweyher2017-si,Caldeweyher2019-hd, Caldeweyher2020-fd}.
The plane wave cutoff energies were 520 eV for the electronic pseudo-wave function and pseudo-charge density, respectively.

\subsection{A2: NNIP Modeling}\label{a2}
In this work, the Allegro model was built through NequIP \cite{Batzner2022-je} that introduced an E(3) symmetry equivariant neural network coded by PyTorch \cite{NEURIPS2019_9015}. 
Our model adopts two layers of 64 tensor features with a $l_\text{max}$ = 2 in full $O(3)$ symmetry. We used a two-body latent multi layer perceptron (MLP) and later latent MLP with hidden dimensions [64, 128, 256, 512] and [512, 512, 512] respectively, both with SiLU nonlinearities. The embedding MLP was a linear projection. For the final edge energy MLP, we used a single hidden layer of dimension 512 and no non-linearity. All four MLPs were initialized according to a uniform distribution of unit variance. Models were trained with a radial cutoff of 5.0 \AA. The loss function was the sum of the per-atom energy, force and stress of the RMSE, and the weight was 1:1:1 respectively. The Adam optimiser, a gradient-based probabilistic algorithm, was used for parameter updating.

\subsection{A3: Free Energy Calculation from Thermodynamic Integration}\label{a3}

In the context of thermodynamic integration, the Helmholtz free energy $F$ difference at a given temperature $T$ between two reference phases (I and II) can be obtained via the integration as follows\cite{Haskins2016-lq,haskins2017finite},
\begin{equation}
\begin{split}
\Delta F_{\text{I} \rightarrow \text{II}} & = \int_0^1 \frac{\partial F}{\partial \lambda} d\lambda \\
& = \int_0^1 \frac{\partial (-k_BT\ln Z)}{\partial \lambda} d\lambda \\
& = -\int_0^1 \frac{k_B T}{Z} \frac{\partial Z}{\partial \lambda} d\lambda \\
& = \int_0^1 \frac{k_B T}{Z} \sum_s \frac{1}{k_BT} \exp[- H(\lambda)/k_BT] \frac{\partial H(\lambda)}{\lambda} d\lambda \\
& = \int_0^1 \bigg\langle\frac{\partial H(\lambda)}{\partial\lambda} \bigg\rangle_\lambda d\lambda,
\end{split}
\end{equation}

where $\lambda$ is a thermodynamic variable along a path that connects I and II, $Z = \sum_s \exp(-H/k_BT)$ is the partition function, $k_B$ is the Boltzmann constant, and $H$ is the Hamiltonian of the system is defined as the sum of the kinetic energy and the potential energy.

In our calculation, we constructed the $\lambda$-path by using the normalized reaction coordinates $\lambda = \cos(\beta)$ from the cell matrix $\boldsymbol{\Omega}$. 
Thus, 
the free energy difference can be numerically evaluated as follows,
\begin{equation}
 \begin{split}
\Delta F_{\text{I} \rightarrow \text{II}}
& = \int_0^1 \bigg\langle \frac{\partial H(\lambda)}{\partial \boldsymbol{\Omega}}\bigg\rangle : \frac{\partial \boldsymbol{\Omega}}{\partial \lambda} d\lambda \\
   &\approx \sum_{k=0}^{n} V_k \langle\boldsymbol{\sigma}_k\rangle  \boldsymbol{\Omega}_k^{-T}: \frac{\partial\boldsymbol{\Omega}_k}{\partial\lambda_k} \Delta \lambda_k \\
 \end{split}\label{eq:A2}
\end{equation}

where $k$ loops over all MD simulations along the path from I to II. In each configuration, both  $V_k$ (the unit cell volume) and $\boldsymbol{\Omega}_k$ (cell matrix) were extracted from the given configuration along the path, and $\boldsymbol{\sigma}_k$ (stress tensors) were obtained by averaging all stress tensors from the MD simulations, and $\partial \boldsymbol{\Omega}_k / \partial \lambda_k$ were obtained from cubic spline interpolation. The ``:" denotes the Frobenius inner product of tensors.

From Eq. \ref{eq:A2}, it is clear that $V_k$, $\boldsymbol{\Omega_k}$ and $\partial \boldsymbol{\Omega}_k / \partial \lambda_k$ are fixed values for a given path. However $\boldsymbol{\sigma}_k$ (in particular, the $\sigma_\text{xz}$ component in the case of aspirin I-II transition) may fluctuate strongly when the configuration is away from the nearby energy minimum, thus leading to a variation of $\Delta F$ value. Hence we use the following relation to evaluate the the error of $\Delta F_{\text{I}\rightarrow \text{II}}$,

\footnotesize
\begin{equation}
   \text{Error}(\Delta F) = \sum_{k=0}^{n-1} \left(\frac{\Delta \lambda_k}{2} \text{RMSE}(f_k)\right)^2 + \sum_{k=0}^{n-1} \left(\frac{\Delta \lambda_k}{2} \text{RMSE}(f_{k+1})\right)^2 \\
\end{equation}
\normalsize

Here, $f_k$ indicates the inside of integration as $f_k=V_k \boldsymbol{\sigma}_k  \boldsymbol{\Omega}_k^{-T}: \frac{\partial\boldsymbol{\Omega}_k}{\partial\lambda_k}$ and RMSE($f$) is root mean square error of $f$ originated in stress tensor fluctuation.

\subsection{A4: Free Energy Calculation from Harmonic Phonons}\label{a4}. 

The vibrational density of states (VDoS) was evaluated to analyze the vibrational differences between aspirin Form I and II as frequency space contributions. To evaluate VDoS, molecular dynamics (MD) simulations with NPT ensembles were performed at 300 K using the NNIP model with 2$\times$4$\times$16 supercells. The VDoS, calculated by Fourier transform from the velocity auto-correlation function of atoms, is shown in Figure \ref{fig:VDOS}. A total of 1000 frames per 25 fs were used as the MD trajectory data. Free energy of harmonic phonon approximation using VDoS spectrum was calculated by following equation\cite{Reilly2014-gp}.
\begin{equation}
   F^{\text{VDoS}} = k_{B}T \int I^{\text{VDoS}}(\omega)  \space\text{ln} \left[ 1- \exp \left( -\frac{\hbar \omega }{k_B T} \right) \right] \\
\end{equation}

Here, $I^{\text{VDoS}}(\omega)$ indicates the intensity of VDoS as $\omega$ is angular frequency of vibration.





\bibliography{ref}

\end{document}